\documentclass{elsart}[10pt]

\usepackage{epsfig}

\usepackage{amssymb}

\begin{document}
\newcommand{\ee}{e^+e^-}
\newcommand{\pee}{\pi^0e^+e^-}
\newcommand{\tp}{\pi^+\pi^-\pi^0}
\newcommand{\pg}{\pi^0\gamma}
\newcommand{\etee}{\eta e^+e^-}
\newcommand{\phph}{\gamma\gamma}
\newcommand{\ompee}{\omega\to\pi^0e^+e^-}
\newcommand{\ometee}{\omega\to\eta e^+e^-}
\newcommand{\rhpee}{\rho\to\pi^0e^+e^-}
\newcommand{\rhetee}{\rho\to\eta e^+e^-}
\newcommand{\ompg}{\omega\to\pi^0\gamma}
\newcommand{\omtp}{\omega\to\pi^+\pi^-\pi^0}
\newcommand{\pc}{\pi^{0}\mathrm{2c}}

\begin{frontmatter}



\title{Study of the $\rho$ and $\omega$ meson decays into 
pseudoscalar meson and  $e^+e^-$ pair with \\the CMD-2 detector}

\author[BINP]{R.R.~Akhmetshin},
\author[BINP,NGU]{V.M.~Aulchenko},
\author[BINP]{V.Sh.~Banzarov},
\author[PITT]{A.~Baratt},
\author[BINP,NGU]{L.M.~Barkov},
%\author[BINP]{S.E.~Baru},
\author[BINP]{N.S.~Bashtovoy},
\author[BINP,NGU]{A.E.~Bondar},
\author[BINP]{D.V.~Bondarev},
\author[BINP]{A.V.~Bragin},
%\author[BINP]{D.V.~Chernyak},
\author[BINP,NGU]{S.I.~Eidelman},
\author[BINP]{D.A.~Epifanov},
\author[BINP,NGU]{G.V.~Fedotovitch},
\author[BINP]{N.I.~Gabyshev},
\author[BINP]{D.A.~Gorbachev},
\author[BINP]{A.A.~Grebeniuk}, 
\author[BINP,NGU]{D.N.~Grigoriev},
%\author[YALE]{V.W.~Hughes}~\footnote{deceased},
\author[BINP]{F.V.~Ignatov},
\author[BINP]{S.V.~Karpov},
\author[BINP,NGU]{V.F.~Kazanin},
\author[BINP,NGU]{B.I.~Khazin},
\author[BINP,NGU]{I.A.~Koop},
\author[BINP,NGU]{P.P.~Krokovny},
%~\footnote{e-mail: krokovny@inp.nsk.su},
%\author[BINP]{L.M.~Kurdadze},
\author[BINP,NGU]{A.S.~Kuzmin},
\author[BINP,BOST]{I.B.~Logashenko},
\author[BINP]{P.A.~Lukin},
\author[BINP]{A.P.~Lysenko},
\author[BINP]{K.Yu.~Mikhailov},
\author[BINP,NGU]{A.I.~Milstein},
\author[BINP,NGU]{I.N.~Nesterenko},
\author[BINP,NGU]{M.A.~Nikulin},
\author[BINP]{V.S.~Okhapkin},
\author[BINP]{A.V.~Otboev},
%\author[BINP,NGU]{A.V.~Pak},
\author[BINP,NGU]{E.A.~Perevedentsev},
\author[BINP]{A.A.~Polunin},
\author[BINP]{A.S.~Popov},
\author[BINP]{S.I.~Redin},
%\author[BOST]{B.L.~Roberts},
\author[BINP]{N.I.~Root},
\author[BINP]{A.A.~Ruban},
\author[BINP]{N.M.~Ryskulov},
\author[BINP]{A.G.~Shamov}, 
\author[BINP]{Yu.M.~Shatunov},
\author[BINP,NGU]{B.A.~Shwartz},
\author[BINP]{A.L.~Sibidanov},
\author[BINP]{V.A.~Sidorov}, 
\author[BINP]{A.N.~Skrinsky},
\author[BINP]{I.G.~Snopkov},
\author[BINP,NGU]{E.P.~Solodov},
%\author[BINP]{P.Yu.~Stepanov},
%\author[BINP]{A.I.~Sukhanov},
\author[PITT]{J.A.~Thompson}\footnote{deceased}, 
\author[BINP]{A.A.~Valishev},
\author[BINP]{Yu.V.~Yudin},
\author[BINP,NGU]{A.S.~Zaitsev},
\author[BINP]{S.G.~Zverev}

\address[BINP]{Budker Institute of Nuclear Physics, 
  Novosibirsk, 630090, Russia}
\address[BOST]{Boston University, Boston, MA 02215, USA}
\address[NGU]{Novosibirsk State University, 
  Novosibirsk, 630090, Russia}
\address[PITT]{University of Pittsburgh, Pittsburgh, PA 15260, USA}

\begin{abstract}
% Text of abstract
Using 3.3 pb$^{-1}$ of data collected with the CMD-2 detector in 
the 720 -- 840 MeV c.m. energy range, the branching fraction of the 
conversion decay $\omega\to\pi^0e^+e^-$ has been measured: 
$\mathcal{B}(\omega\to\pi^0e^+e^-)=(8.19\pm0.71\pm0.62)\cdot10^{-4}$.
The upper limits for the branching fractions of the following
conversion decays have been obtained at the 90\% confidence level: 
$\mathcal{B}(\rho\to\pi^0e^+e^-)<1.6\cdot10^{-5}$,
$\mathcal{B}(\rho\to\eta e^+e^-)<0.7\cdot10^{-5}$ and
$\mathcal{B}(\omega\to\eta e^+e^-)<1.1\cdot10^{-5}$.

\end{abstract}

\end{frontmatter}

% main text
\section{Introduction}
\label{s1}
Measurement of branching fractions and transition form factors of
conversion decays provides an important test of vector dominance
model~\cite{Bauer,Donnel} and an accurate background estimation 
in searches for quark-gluon plasma involving a lepton 
pair~\cite{Shuryak,QGPexceed}.

The expected branching fractions for the 
$\rho$ and $\omega$ conversion decays are
$\mathcal{B}(\ompee)=(7.2-8.0)\cdot 10^{-4}$,
$\mathcal{B}(\rhpee)=(4.1-6.5)\cdot 10^{-6}$ and an order of magnitude 
smaller for decays with muons and/or $\eta$ meson in the final
state~\cite{Landsberg,Eidelman,Faessler}.
Various conversion decays of the $\phi$ meson were studied at
CMD-2~\cite{phietaee,phipi0ee} and SND~\cite{sndconv,sndphipi0ee}. 
For the $\rho$ and $\omega$ mesons experimental information is rather 
scarce: only the branching fractions for the 
decays $\ompee$ and $\omega\to\pi^0\mu^+\mu^-$ were measured in 
experiments at ND~\cite{ND} and Lepton-G~\cite{Lepton-G} with an 
accuracy $\sim 30\%$.

Conversion decays of a vector meson $V$ into a pseudoscalar meson 
$P$ and lepton pair $l^+l^-$ ($V\to P l^+l^-$) are closely related to 
the corresponding radiative meson decays $V\to P\gamma$~\cite{Landsberg}
%The lepton pair in conversion decays is produced via a virtual photon 
%$\gamma^{*}$ with mass $q=M_{inv}(l^+l^-)$. 
and provide a possibility to study
a transition $V\to P$ form factor, $F_{VP}(q^2)$, as a function of
squared mass of virtual photon,  $q^2=M^2_{inv}(l^+l^-)$.

In this work we present the result of the studies of the $\rho$ and 
$\omega$ conversion decays into a $e^+e^-$ pair and pseudoscalar 
meson ($\pi$ or $\eta$) performed with the CMD-2 detector at the 
VEPP-2M collider~\cite{theVEPP}.
The analysis is based on a data sample collected at 19 energy points 
in the 720~--~840~MeV c.m. energy range and corresponding to 
3.3~pb$^{-1}$ of integrated luminosity. This statistics contains
about 3.3$\cdot$10$^6$ $\rho$ and 1.8$\cdot$10$^6$ $\omega$
decays.
More detailed description of this analysis can be found in~\cite{pre}.

% \appendix
% \section{}
% \label{}
\section{Experiment}
\label{s2}
The general purpose detector CMD-2 has been described in 
detail elsewhere~\cite{thedet}. The tracking system consists of the
cylindrical drift chamber (DC) with 250~$\mu$ resolution transverse 
to the beam plane and double-layer multiwire proportional 
Z-chamber, both also used for the trigger. The tracking system is 
placed inside a thin 
(0.38~X$_0$) superconducting solenoid with a field of 1~T.
The barrel CsI calorimeter with a thickness of 8.1~X$_0$ placed
outside  the solenoid has energy resolution for photons of about
9\% in the energy range from 100 to 700~MeV. The angular resolution is 
of the order of 0.02 radians. The end-cap BGO calorimeter with a 
thickness of 13.4~X$_0$ placed inside the solenoid 
has energy and angular resolution varying from 9\% to 4\% and from 
0.03 to 0.02 radians, respectively, for the photon energy in the range 
100 to 700~MeV. The barrel and end-cap calorimeter systems cover a 
solid angle of $0.92\times4\pi$ radians. 

\section{Data analysis}
The decay $\ompee$ has been studied using the $\pi^0$
dominant decay mode $\pi^0\to\phph$. It corresponds to a final 
state with two opposite charge particles and two photons. 
%The effective mass $M_{inv}(\phph)$ of two photons should be
%equal to $\pi^0$ mass within the detector resolution.

One of the significant resonant backgrounds comes from the 
$\omtp$ decay which has the same
topology of the final state and more than three orders of magnitude larger
probability. Another source of resonant background is the $\ompg$ decay 
followed by the Dalitz decay of the $\pi^0$ or $\gamma$-quantum 
conversion in the material in front 
of the drift chamber. Most conversions occur in the beam pipe, at 
the distance of 1.8 cm from the interaction point. Since the DC
spatial resolution is not sufficient to separate events with
conversions in the beam pipe from those where it occurs at the interaction
point, the contribution of this background was subtracted based on Monte Carlo 
simulation (MC). The non-resonant background includes contributions from 
the following QED processes with the same final state topology: 
$\ee\to\ee\phph$, $\ee\to3\gamma$ followed by $\gamma$-quantum 
conversions, $\ee\to\ee\gamma$ with one background
photon as well as two-quantum annihilation followed by a
$\gamma$-quantum conversion and one
background photon in calorimeters.

The data analysis consists of two parts: a measurement of 
the branching fraction and a study of the transition form factor. 
The selection criteria used for the measurement of the branching fractions
were tuned to select a ``pure'' set of events under study and thus
reject events with a large invariant mass of $e^+e^-$ which contain a
lot of background. However, this very range is of main interest for
a study of the transition form factor. Therefore, for the latter a 
special set of selection criteria including a technique of e/$\pi$ 
separation~\cite{phipi0ee,4pi} was
applied to suppress the main background from 3$\pi$ events.
% These criteria Firstly, the analysis on
%the measurements of branching fraction value is presented. The 
%selection criteria used in this analysis were tuned to select a 
%``pure'' set of events under study. These criteria
%could not be used in the analysis of the transition form factor 
%since they reject events with a large invariant mass of
%$e^+e^-$ which are of interest. To suppress the main background 
%from 3$\pi$ decay in the analysis of the transition form factor, the 
%universal technique of e/$\pi$ separation~\cite{phipi0ee,4pi} was used.

\subsection{Selection of $\rho(\omega)\to\pi^0\ee$ events}
\label{ss31}
\begin{figure}[tbh]
\centerline{\psfig{file=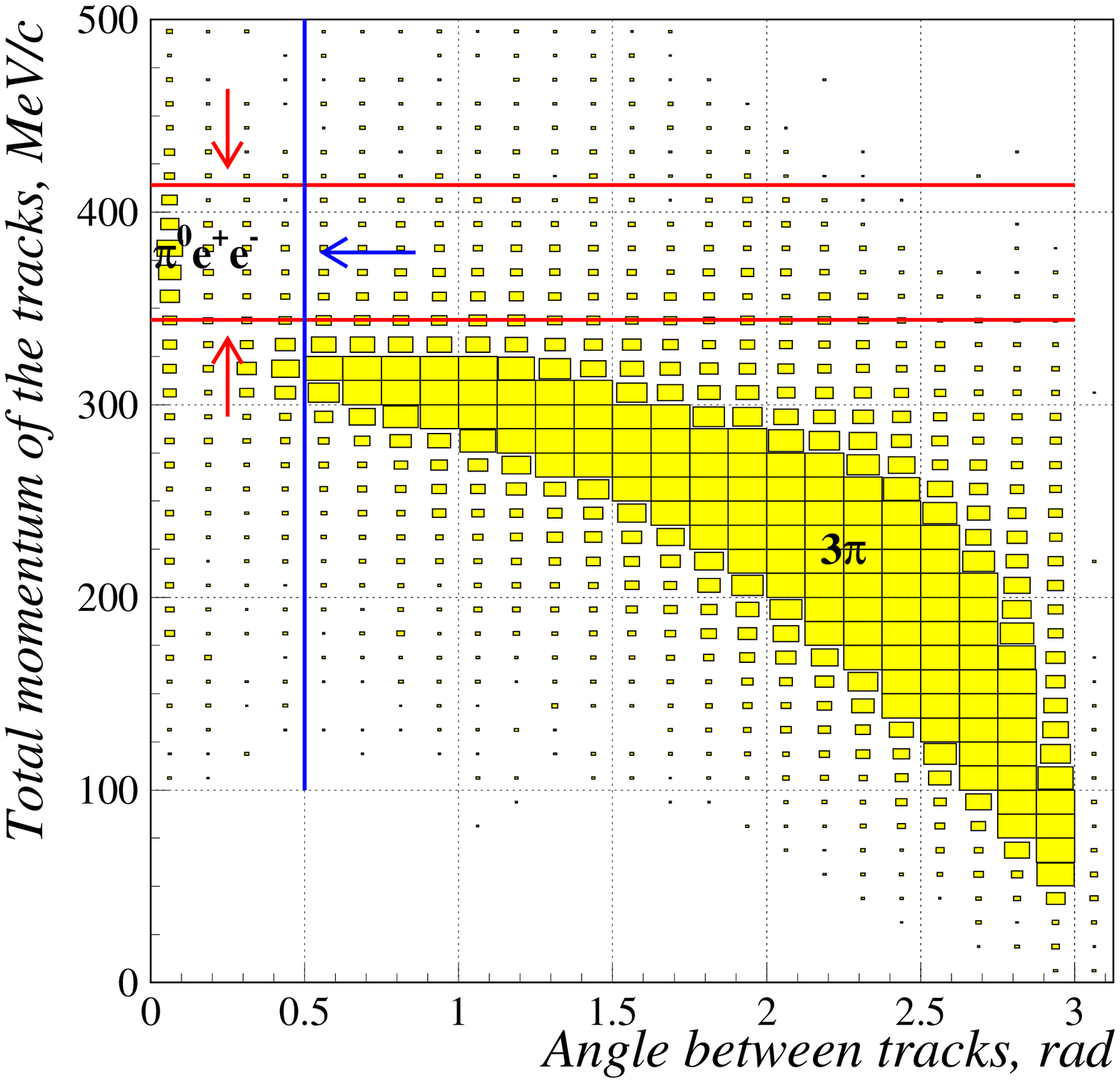,width=0.47\textwidth}\hfill\psfig{file=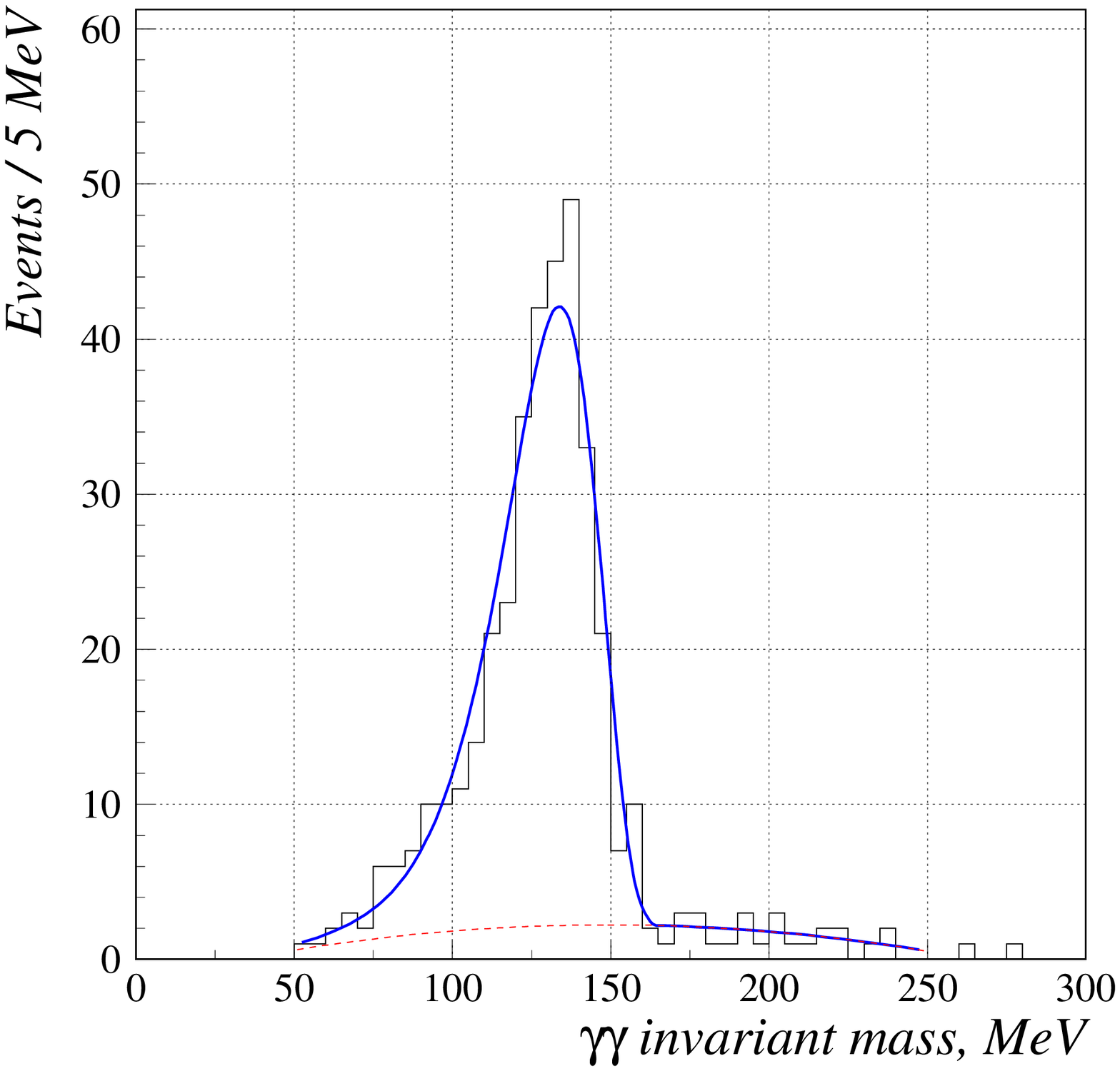,width=0.47\textwidth}}
%\vspace*{8pt}
\parbox[t]{0.47\textwidth}{\caption{\it
Opening angle of two tracks $\Delta\psi$ versus 
their total momentum $p$. The horizontal lines show the
    cut on the total momentum and the vertical one shows 
the cut on the opening angle of the tracks.
  }\label{f:mompar}}
\hfill
\parbox[t]{0.47\textwidth}{\caption{\it
    The $M_{inv}(\phph)$ distribution (histogram). The results of the fit
    are shown with a solid (signal) and dotted (background) line.
  }\label{f:invm}}
\end{figure}

At the first stage of selection, the following criteria were applied 
to events with two tracks and at least two photons to
enrich a data sample with events of the studied decays: 
\begin{itemize}
\item the photon energy $E_{\gamma}>40$~MeV and its polar angle
  $0.5<\theta_{\gamma}<\pi-0.5$ to suppress background photons in 
the calorimeters;
\item the impact parameter of the tracks $\rho<1$~cm and Z-coordinate of 
the vertex $|Z_{vert}|<5$~cm to reject cosmic rays and beam background 
events;
\item the total momentum of the tracks $p=|\vec{P}_1+\vec{P}_2|$ does not 
strongly differ from the photon momentum $p_{\gamma}$ in the $\ompg$
decay at a given
energy $|p-p_{\gamma}|<35$~MeV/c to suppress $\omtp$ events as well
as $\ompg$ events followed by the Dalitz decay of the $\pi^0$.
This condition is illustrated with horizontal lines in 
Fig.~\ref{f:mompar} where events satisfying previous requirements
are shown;
\item the angle between the total momentum of the tracks and each photon
is more than 1.7 to suppress QED events;
\item the invariant mass of the electron-positron pair and most 
energetic photon $M_{inv}(\ee\gamma_1)$ is less than 
$1.9\cdot E_{beam}$ to suppress  $\ee\to\phph$  events followed by a 
conversion of one $\gamma$-quantum;
\item the opening angle of the tracks $\Delta\psi<0.5$ to suppress 
events of the $\omtp$ decay, this condition is shown in 
Fig.~\ref{f:mompar} with a  vertical line.
\end{itemize}
Using these criteria 390 events were selected.

%The events of decays under study have two photons in final
%state originated from $\pi^0$. 
At the next stage,  a fit of the $\phph$ invariant mass distribution
($M_{inv}(\phph)$) was performed at each energy point to determine 
the number of events with $\pi^0$ in the final state denoted 
hereafter as ``$\pc$'' ($\ompee$, $\ompg$ followed by a 
$\gamma$-conversion and $\omtp$).
%at each energy point a fit of the $\phph$ invariant mass distribution
%($M_{inv}(\phph)$) was performed.
%The events of studied decays have two pions originated from 
%$\pi^0$ in final state. The determination of events with $\pi^0$ in 
%final state, so called signal event, was made from analysis of two 
%gamma invariant mass spectra $M_{\gamma\gamma}$ of selected events.
%At each energy point the $M_{inv}(\phph)$ distribution 
%was fitted to determine the number of signal events in this data
%sample.
The parameterization included a logarithmic Gaussian for the signal
and a second order polynomial for the background. The shape of these 
functions was fixed, so the floating parameters were the numbers of
events of each type only. The shape of background was taken from 
a study of the $M_{inv}(\phph)$ distributions from the $\omega$ decays 
into $\pg$ and $\tp$ specially selected in the whole energy range. 
The shape of the signal was obtained from the
fit of the $M_{inv}(\phph)$ distribution
over all selected events with floating parameters of the function 
describing the signal, see Fig.~\ref{f:invm}. The total number
of selected $\pc$ events is 316.
%The $M_{inv}(\phph)$ distribution over all selected events was
%used to determine the total number of signal events as well as 
%the shape of the signal. It was done by the fit of this distribution 
%with a sum of two functions that describe the response of signal 
%(logarithmic gaussian) and background (second order polynomial) in 
%this distribution (Fig.~\ref{f:invm}).
%The shape of background was taken from the study of $M_{inv}(\phph)$
%distributions from $\omega$ decays into $\pg$ and $\tp$ events. The 
%$M_{inv}(\phph)$ distributions fit with the sum of signal and 
%background was used to obtain the number of signal and background
%events in each energy point. The shape of functions describing
%the signal and background in the $M_{inv}(\phph)$ distribution were 
%fixed, the varied paremeters were the number of events of each type only.

\subsection{Study of main background processes}
\begin{figure}[tbh]
  \centering
  \psfig{file=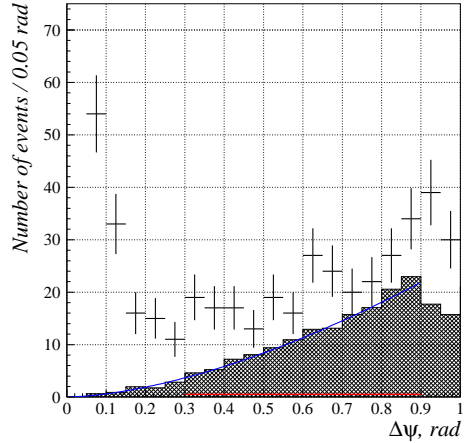,width=0.47\textwidth}
  %\vspace*{8pt}
  \caption{\it
    The $\Delta\psi$ distribution for the data  
(crosses) and $\omtp$ simulation events (hatched histogram). All 
selection criteria but that on $\Delta\psi$ were applied. The number 
of $\tp$ MC events was approximated with a function
$x^{\alpha}$ shown with a solid line.}
  \label{f:dpsi}
\end{figure}

At this stage of analysis the expected number of background events
was evaluated using the $\Delta\psi$  distribution. Its difference 
for events of the process under study and those of 
background was employed in the analysis: the former are strongly 
peaked near zero while events coming from the $\omtp$ decay have a 
smooth wide distribution. Events of the $\ompg$ decay with conversion,
which are also strongly peaked near zero, predominantly populate 
the angular range $\Delta\psi < 0.3$.
% was determined from the
%track opening angle distribution ($\Delta\psi$). The $\Delta\psi$ 
%$\Delta\psi$  distribution. 
The  $\Delta\psi$ distribution for the data after imposing all 
selection conditions but that on $\Delta\psi$ is shown by crosses in 
Fig.~\ref{f:dpsi}. Using the modified requirement 
$0.3 < \Delta\psi < 0.9$ we reject most of the events coming from the 
$\ompg$ decay with conversion and keep events 
under study as well as background events from  the $\omtp$ and QED.
The MC $\Delta\psi$ distribution for the main background of $\omtp$ 
shown with a hatched histogram in Fig.~\ref{f:dpsi} was approximated
with a function $f(x)\sim x^{\alpha}$ with $\alpha=1.63\pm0.10$  
in the $\Delta\psi<0.9$ angular range.
%Using the modified requirement $0.3 < \Delta\psi < 0.9$ we reject
%events coming from the $\ompg$ decay with conversion keeping events 
%under study and background events from the
%$\omtp$ and QED.
% so the selection cut $0.3<\Delta\psi<0.9$ was applied to the data
%for analysis of background coming from events of $\omtp$ decays.
The energy dependence of the number of events selected at the first stage 
(with modified requirements on $\Delta\psi$) was 
used to determine the number of $\omtp$ events.

%The dependence of number of the selected events on $s$ was used to 
%determine the number of $\omtp$ events. 
To this end, the number of selected events at each point was written as
\begin{equation}
  N_i = \bigl(\sigma_{\mathrm{res}}(s_i)\,(1+\delta_i)+\sigma_{QED}(s_i)\bigr)\cdot L_i,
  \label{m:cs3pi}
\end{equation}
where $\sigma_{QED}(s)=\sigma^0_{QED}\cdot m_\omega^2/s$,
$\sigma_{\mathrm{res}}(s)$ is the cross section of the $\omega$ meson production
at the squared c.m. energy $s$,
$L_i$ denotes the integrated luminosity at the $i$-th energy point, 
and $\delta_i$ is a radiative correction. The floating parameters 
are $\sigma^0_{QED}$ and $\sigma^0_{\mathrm{res}}$ --- cross section values at 
the resonance peak. The resulting number of resonance events was 
calculated as a sum over all energy points: 
$N_{res}=\sum\sigma_{\mathrm{res}}(s_i)\,L_i\,(1+\delta_i)=210\pm17$.
After subtraction of the  $\ompee$  contribution, the number of
$\tp$ events $N_{\tp}=180\pm20$ in the $0.3<\Delta\psi<0.9$ range.
The above description of the $\Delta\psi$ distribution was used to 
calculate the number of $\omtp$ events in the $\Delta\psi<0.5$ range:
\begin{equation}
N_{3\pi}=40.6\pm6.4\pm4.6.
\label{m:n3pi}
\end{equation}
The systematic error comes from the  uncertainty of the $\Delta\psi$  
distribution approximations and that of the subtraction of the 
$\ompee$  contribution.
Since events of $\ompg$ decays could not be distinguished from those 
under study, the detection efficiency for the former was determined as
$\varepsilon^{\pee}_{\mathrm{det}}\cdot P_{\mathrm{conv}}$, where
$\varepsilon^{\pee}_{\mathrm{det}}$ denotes the detection efficiency for
$\pee$ events and $P_{\mathrm{conv}}$ is a probability for a
monoenergetic photon from the $\ompg$ decay to convert in the 
material in front of DC.
The $P_{\mathrm{conv}}$ value was obtained from simulation:
\begin{equation}
P_{\mathrm{conv}}=(1.91\pm0.06)\cdot10^{-3}.
\label{m:pconv}
\end{equation}
The error of the $P_{\mathrm{conv}}$ value above comes from
the uncertainties in a thickness and composition of materials in front 
of DC.
\subsection{Approximation of $\pc$ events}
%The signal events come from $\omega\to\pee,\,\tp,\,\pg$ decays 
%and probably QED events.
At the last stage of analysis, the energy dependence of the number of 
$\pc$ events is fitted
%In this fit the energy dependence of the number of $\pc$ events
with a sum of the contributions from the
$\omega\to\pee$, $\tp$, $\pg$ decays and possibly remaining QED events:
\begin{equation}
  N_{\pc,\,i}=N_{\pee,\,i}+N_{\pg,\,i}+N_{3\pi,\,i}+N_{QED,\,i}.
  \label{m:npi02c}
\end{equation}
The number of $\pee$ events at the $i$-th energy point was described
by the expression (\ref{m:nee}):
\begin{equation}
  N_{\pee,\,i}=
  \sigma_{\pee}(s_i)\,L_i\,(1+\delta_i)\,
  \varepsilon^{\pee}_{\mathrm{det},\,i}\,
  \varepsilon_{\Delta\psi,\,i}\,\varepsilon_{\mathrm{trig},\,i}\,
  \mathcal{B}(\pi^0\to\gamma\gamma),
\label{m:nee}
\end{equation}
where $L_i$ is an integrated luminosity, $\delta_i$ is a radiative 
correction, $\varepsilon^{\pee}_{\mathrm{det},\,i}$,
$\varepsilon_{\Delta\psi,\,i}$, $\varepsilon_{\mathrm{trig},\,i}$ denote
the detection efficiency, efficiency of the reconstruction of close 
tracks and trigger efficiency at the $i$-th energy point, respectively. The 
energy dependence of the Born cross section
$\sigma_{\pi^0\mathrm{e}^+\mathrm{e}^-}(s)$ was
written using the relativistic Breit-Wigner approach with the $\rho$ 
and $\omega$ meson contributions:
\begin{equation}
\sigma_{\pee}(s) = \frac{q^3(s)}{s^{3/2}}\cdot
\bigl|A_{\rho}(s)+A_{\omega}(s)+a_0\bigr|^2.
\label{m:bw}
\end{equation}
Here $q(s)$ is a phase space factor:
\begin{equation}
  q(s)=\frac{\sqrt{s}}{2}\,\Bigl(1-\frac{m^2_{\pi^0}}{s}\Bigr),
  \label{m:phase}
\end{equation}
$A_V(s)$ is an amplitude of the vector meson $V$:
\begin{equation}
A_V(s)=\frac{m_V^2\,\Gamma_V\,\sqrt{\sigma^0_V}\,f_V}{D_V(s)\,q^{3/2}(m^2_V)},
%             A_V(s)=\frac{m_V\Gamma_Vf_V(s)}{m_V^2-s-i\sqrt{s}\Gamma_V(s)}
%             \cdot\sqrt{\frac{m_V^3}{q(m_V^2)^3}\sigma_V^0}.
\label{m:av}
\end{equation}
and the additional constant $a_0$ in the amplitude describes a 
possible contribution of higher resonances.
The quantities $m_V$ and  $\sigma^0_V$ are the vector meson mass and 
the cross section at its peak, respectively,  with $\sigma^0_V$
calculated without taking into account other contributions, and
$f_V=e^{i\phi_V}$ is a phase factor. To describe the $\rho$--$\omega$ 
interference, the model with energy-independent interference phases
was used with $\phi_\omega$ set to zero and 
$\phi_{\rho}=-13^0$ \cite{pi0g}. The $1/D_V(s)$ is a vector meson 
propagator described by the expression:
\begin{equation}
  D_V(s) = m_V^2-s-im_V\Gamma_V(s).
  \label{m:vprop}
\end{equation}
The quantities $\Gamma_V(s)$ and $\Gamma_V=\Gamma_V(m_V^2)$ are
vector meson width at the squared c.m. energy $s$ and at the vector 
meson mass, respectively. 

The number of $\pg$ events was written as
\begin{equation}
  N_{\pg,\,i}=
  \sigma_{\pg}(s_i)\,L_i\,(1+\delta^{\pg}_i)\,
  \varepsilon^{\pee}_{\mathrm{det},\,i}\,P_{\mathrm{conv}}\,
  \varepsilon_{\Delta\psi,\,i}\,\varepsilon_{\mathrm{trig},\,i}\,
  \mathcal{B}(\pi^0\to\gamma\gamma),
\label{m:pg}
\end{equation}
with $\sigma_{\pg}(s)$ taken from~\cite{pi0g} and
 a photon conversion probability $P_{conv}$ taken from (\ref{m:pconv}).
To describe the energy dependence of $\tp$ events, the relativistic 
Breit-Wigner with the $\omega$ meson contribution was used:
\begin{equation}
  N_{3\pi,\,i}=\sigma_{\mathrm{BW}}(s_i)\,L_i\,(1+\delta^{3\pi}_i)\,\varepsilon_{3\pi}.
  \label{m:3pi}
\end{equation}
Since the total number of $\tp$ events was previously found, see
expression~(\ref{m:n3pi}),
the normalization factor $\varepsilon_{3\pi}$ was determined from the sum
$N_{3\pi}=\sum\sigma_{\mathrm{BW}}(s_i)\,L_i\,(1+\delta^{3\pi}_i)\,\varepsilon_{3\pi}$.
A possible contribution of QED events which can, e.g.,  appear because
of the incorrect background subtraction was evaluated from the following 
expression:
\begin{equation}
  N_{QED,\ i}=\sigma^0_{QED}\,L_i\frac{m_{\omega}^2}{s_i}.
  \label{m:qed}
\end{equation}

The detection efficiencies $\varepsilon^{\pee}_{\mathrm{det},\,i}$ were 
determined using the Monte-Carlo simulation
taking into account initial and final state radiation. The calculation of
radiative corrections $\delta$ followed Ref.~\cite{rc}.

The events under study can be triggered by two
independent triggers: charged and neutral ones.
The overall trigger efficiency is estimated to be 99\%.
The estimation of this value has been
performed by analysis of the trigger signals in selected events.

Since Monte-Carlo simulation does not completely describe the
experiment, a correction  $\varepsilon_{\Delta\psi}$ for a  difference 
between the  efficiencies of close track reconstruction in simulation 
and experiment was included to describe the data.
Its value was obtained using events of $\omtp$ decays
followed by the conversion decay $\pi^0\to\ee\gamma$ with a similar
$\Delta\psi$ distribution. The integrated luminosity was
determined using events of large angle Bhabha scattering
with radiative corrections taken into account according to \cite{Arbuzov}.  

Table \ref{t:data} shows the detailed information about experiment 
including efficiencies of detection, trigger and close track 
reconstruction together with the number of $\pc$ events and visible 
cross section calculated from the following expression: 
\begin{equation}
  \sigma^{vis}_{\pc} = \frac{N_{\pc}^{\mathrm{exp}}}{L\,(1+\delta)\,
    \varepsilon^{\pee}_{\mathrm{det}}\,\varepsilon_{\Delta\psi}\,
    \varepsilon_{\mathrm{trig}}\,\mathcal{B}(\pi^0\to\gamma\gamma)}.
  \label{m:csvis}
\end{equation}

\begin{table}[tbh]
\caption{The energy, integrated luminosity, radiative correction, 
detection efficiency, the reconstruction efficiency of close tracks, 
trigger efficiency, number of $\pc$ events and visible cross section of
$\pc$ events.}
\vspace*{1mm}
\begin{tabular}{cccccccc} \hline \hline
$\sqrt{s}$, MeV & $L$, nb$^{-1}$ & $\delta$ & $\varepsilon^{\pee}_{\mathrm{det}}$ &
$\varepsilon_{\Delta\psi}$ & $\varepsilon_{\mathrm{trig}}$ &
$N^{\mathrm{exp}}_{\pc}$ & $\sigma^{vis}_{\pc}$, nb \\ \hline
720 & 183.86 & -0.099 & 0.149 & 0.906 & 0.989 & 0.8$_{-0.6}^{+1.2}$ & 0.04$_{-0.03}^{+0.06}$\\
750 & 148.66 & -0.119 & 0.149 & 0.906 & 0.989 & 1.5$_{-0.9}^{+1.6}$ & 0.09$_{-0.06}^{+0.10}$\\
760 & 157.11 & -0.152 & 0.149 & 0.906 & 0.989 & 3.4$_{-1.5}^{+2.2}$ & 0.20$_{-0.09}^{+0.13}$\\
770 &  45.81 & -0.186 & 0.149 & 0.906 & 0.989 & 0.0$_{-0.0}^{+1.0}$ & 0.00$_{-0.00}^{+0.21}$\\
774 & 119.57 & -0.211 & 0.148 & 0.906 & 0.989 & 3.1$_{-1.5}^{+2.1}$ & 0.26$_{-0.12}^{+0.17}$\\
778 & 128.25 & -0.218 & 0.148 & 0.906 & 0.989 & 19.4$_{-4.4}^{+5.0}$ & 1.50$_{-0.34}^{+0.39}$\\
780 & 132.82 & -0.229 & 0.148 & 0.906 & 0.989 & 21.4$_{-4.9}^{+5.1}$ & 1.61$_{-0.37}^{+0.38}$\\
781 & 187.09 & -0.224 & 0.148 & 0.906 & 0.989 & 28.8$_{-5.3}^{+5.9}$ & 1.52$_{-0.28}^{+0.31}$\\
782 & 240.98 & -0.211 & 0.148 & 0.906 & 0.989 & 56.8$_{-7.9}^{+8.5}$ & 2.30$_{-0.32}^{+0.34}$\\
783 & 187.34 & -0.191 & 0.148 & 0.906 & 0.989 & 54.1$_{-7.6}^{+8.2}$ & 2.77$_{-0.39}^{+0.42}$\\
784 & 237.05 & -0.167 & 0.149 & 0.906 & 0.995 & 56.3$_{-7.9}^{+8.4}$ & 2.20$_{-0.31}^{+0.33}$\\
785 & 191.51 & -0.138 & 0.149 & 0.906 & 0.998 & 35.0$_{-5.6}^{+7.5}$ & 1.63$_{-0.26}^{+0.35}$\\
786 & 108.63 & -0.112 & 0.149 & 0.906 & 0.998 & 9.2$_{-2.7}^{+3.4}$ & 0.71$_{-0.21}^{+0.26}$\\
790 & 117.03 & 0.053 & 0.149 & 0.906 & 0.998 & 9.5$_{-2.8}^{+3.4}$ & 0.60$_{-0.17}^{+0.22}$\\
794 & 121.81 & 0.260 & 0.149 & 0.906 & 0.998 & 5.1$_{-1.9}^{+2.6}$ & 0.27$_{-0.10}^{+0.14}$\\
800 & 195.50 & 0.427 & 0.150 & 0.907 & 0.998 & 2.6$_{-1.3}^{+2.0}$ & 0.07$_{-0.04}^{+0.06}$\\
810 & 181.17 & 0.828 & 0.151 & 0.907 & 0.998 & 2.6$_{-1.3}^{+2.0}$ & 0.07$_{-0.03}^{+0.05}$\\
820 & 185.52 & 1.023 & 0.151 & 0.907 & 0.998 & 2.7$_{-1.3}^{+2.0}$ & 0.06$_{-0.03}^{+0.05}$\\
840 & 457.87 & 1.260 & 0.151 & 0.907 & 0.998 & 3.2$_{-1.5}^{+2.1}$ & 0.03$_{-0.01}^{+0.02}$\\ \hline \hline
\end{tabular}
\label{t:data}
\end{table}

The energy dependence of the number of $\pc$ events was fitted 
using the maximum likelihood method. The minimization function is 
shown here:
\begin{equation}
\mathcal{L}=\sum_{i=1}^{n}\Bigl(\frac{(N_{\mathrm{\pc},\,i}-N_{\mathrm{\pc},\,i}^{\mathrm{exp}})^2}{\sigma_{\pm,\,i}^2}\Bigr)+
\frac{(\tilde{N}_{3\pi}-N_{3\pi})^2}{N_{3\pi}}+
\frac{(\tilde{m}_\omega-m_{\omega})^2}{\varepsilon(m_\omega)^2}+
\frac{(\tilde{\Gamma}_\omega-\Gamma_{\omega})^2}{\varepsilon(\Gamma_\omega)^2},
\label{m:likelihood}
\end{equation}
where $N_{\mathrm{\pc},\,i}=N_{\pee,\,i}+N_{\pg,\,i}+N_{3\pi,\,i}+N_{QED,\,i}$ and
$N_{\mathrm{\pc},\,i}^{\mathrm{exp}}$ is the experimental
number of $\pc$ events at the $i$-th energy point. The quantities 
$\sigma^2_{\pm,\,i}$ are asymmetric variances of the number of $\pc$ events.
The floating parameters are two branching fractions
$\mathcal{B}(\ompee)$ and $\mathcal{B}(\rhpee)$, and the constants
$a_0$ and $\sigma^0_{QED}$. Since the statistics is not sufficient to
determine precisely main parameters of the $\omega$ meson,
the fit allows  $\omega$ meson mass $\tilde{m}_\omega$,
width $\tilde{\Gamma}_\omega$ and the number of $\tp$ events 
$\tilde{N}_{3\pi}$
(three last terms in expression~(\ref{m:likelihood})) to float around
their expected values within their estimated uncertainties.
%The expected value and estimated error was used in fit for the
%following parameters: $\omega$ meson mass $\tilde{m}_\omega$,
%width $\tilde{\Gamma}_\omega$ and the number of $\tp$ events 
%$\tilde{N}_{3\pi}$ (three last terms in 
%expression~(\ref{m:likelihood})). 
The central value and error
$m_\omega$ and $\Gamma_\omega$ were taken from PDG~\cite{PDG2004}
while those for ${N}_{3\pi}$ were taken from~(\ref{m:n3pi}).
The $\rho$ meson parameters were fixed from the work \cite{reanal}.
\begin{table}[tbh]
\caption{The fit results in various models. Only statistical errors 
are shown.}
\vspace*{1mm}
\begin{tabular}{rccccc} \hline \hline
    Model & $\mathcal{B}$($\omega$), 10$^{-4}$ & 
$\mathcal{B}$($\rho$), 10$^{-4}$ &
    $\sigma_{QED}$, pb
    & $a_0$, nb$^{1/2}$ & $\chi^2$/n.d.f. \\ \hline
    I & 8.19$\pm$0.81 & fixed & -0.1$\pm$2.5 &
    $\equiv$ 0 & 19.11/17 \\
    II & 8.20$\pm$0.77 & fixed & $\equiv$ 0 &
    0.10$\pm$0.14 & 18.17/17 \\
    III & 8.19$\pm$0.71 & fixed & $\equiv$ 0 &
    $\equiv$ 0 & 19.11/18 \\ 
    IV & 8.33$\pm$1.37 & 0.032$\pm$0.081 & 
    $\equiv$ 0 & $\equiv$ 0 & 19.06/17 \\ \hline \hline
\end{tabular}
\label{t:fit}
\end{table}

Four different models were used. 
The ratio of $\sigma^0_{\omega}$ and $\sigma^0_{\rho}$ values,
cross sections at the $\rho$ and $\omega$ meson peak, respectively, was
fixed in some models from the results of the study of the process 
$\ee\to\pg$~\cite{pi0g}, so that
\begin{equation}
\frac{\sigma^0(\rho\to\pee)}{\sigma^0(\omega\to\pee)}\approx
\frac{\sigma^0(\rho\to\pg)}{\sigma^0(\omega\to\pg)}=
(3.7\pm1.0)\cdot 10^{-3}.
\label{m:s0rhoom}
\end{equation}
The theoretical precision of the approximate equality in (\ref{m:s0rhoom})
is evaluated as 3\% for the applied selection criteria because
of small $q^2$ in selected events. The expression (\ref{m:s0rhoom}) fixes
the ratio between the branching fractions
$\mathcal{B}(\rhpee)/\mathcal{B}(\ompee)=(1.6\pm0.4)\cdot10^{-3}$.

\begin{figure}[tbh]
\centerline{
  \psfig{file=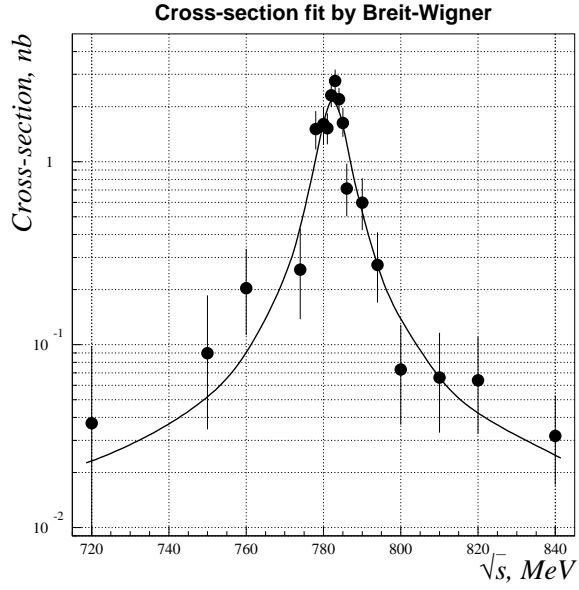,width=0.6\textwidth}
  }
%\vspace*{8pt}
\caption{\it Visible cross section of $\pc$ events. The fit in model III
  is shown.\label{f:sigbw}}
\end{figure}

A fit in the first model was performed with floating
$\sigma^0_{QED}$ to evaluate a possible contribution of QED events. 
As a result, the contribution of QED to $\pc$ events is negligible,
so in the other models this parameter was fixed to zero.
Later, the fit in model II had a floating $a_0$ to evaluate a possible 
contribution from higher resonances. As a result, this
contribution $a_0$ does not differ from zero within a statistical error.
In model III a fit with a fixed ratio $\sigma^0_\rho/\sigma^0_\omega$ was performed 
to obtain a branching fraction for the $\ompee$ decay
\begin{equation}
  \mathcal{B}(\ompee)=(8.19\pm0.71\pm0.62)\cdot10^{-4}.
  \label{m:ompeeval}
\end{equation}
The graphical demonstration of this fit together with the visible cross section of
the $\pc$ events is shown in Figure~\ref{f:sigbw}.
As a result of this fit the number of $\ompee$ events was determined to be 232. 
Finally, a fit in model IV with 
floating $\mathcal{B}(\ompee)$ and $\mathcal{B}(\rhpee)$ was performed to set an 
upper limit on the branching fraction for the $\rhpee$ decay:
\begin{equation}
  \mathcal{B}(\rhpee)<1.6\cdot10^{-5}\ (90\%\ \mathrm{C.L.}).
  \label{m:rhpeelim}
\end{equation}

The obtained value of the $\ompee$ branching fraction 
(expression~(\ref{m:ompeeval}))
agrees with the previous measurement from ND~\cite{ND} but has two times
better accuracy. It is also consistent with the theoretical
predictions~\cite{Eidelman,Faessler}.

\subsection{Systematic errors}
The main sources of systematic uncertainties of the branching
fractions are listed in~Table~\ref{t:syserr}.
\begin{table}[tbh]
  \caption{The main sources of systematic uncertainties}
  \vspace*{1mm}
  \begin{tabular}{lc} \hline \hline
    Source & Uncertainty, \% \\ \hline
Reconstruction efficiency of close tracks $\varepsilon_{\Delta\psi}$ &
 4.7 \\
    Background subtraction & 3.6 \\
    Detection efficiency $\varepsilon^{\pee}_{\mathrm{det}}$ & 2.2 \\ 
    Trigger efficiency $\varepsilon_{\mathrm{trig}}$ & 2.0 \\
    Parameters of $\rho$ and $\omega$  & 2.0 \\
    Mixing parameter $a=\sigma^0_{\rho}/\sigma^0_{\omega}$ & 1.9 \\ 
    Integrated luminosity & 1.4 \\
    Form factor model  & 1.2 \\
    Radiative corrections  & 1.2 \\
    Probability of conversion  & 1.0 \\ \hline
    Total &  7.6 \\ \hline \hline
  \end{tabular}
  \label{t:syserr}
\end{table}
A contribution of each parameter to the total systematic error was 
estimated as a change of the branching fraction when this parameter 
was varied within its  measurement uncertainty. 
The main contribution comes from the reconstruction efficiency of close
tracks and it is completely determined by statistics of the test
samples of the $\omega\to\tp$ decays followed by $\pi^0\to\ee\gamma$. 
The background subtraction error originates from the inexact knowledge 
of the shape of the $M(\gamma\gamma)$ distribution for background
events  and from the error in the number of $\tp$ events.
%The background subtraction is an error arising from
%the shape uncertainty of background events in the $M(\gamma\gamma)$ 
%spectrum of selected events and shape uncertainties of selected 
%$\tp$ events in tracks opening angle $\Delta\psi$ spectrum.
Since the reconstruction of close tracks is sensitive to
the presence of the charged trigger, a  contribution of the trigger 
efficiency  was conservatively
evaluated as a change of the branching fraction when we additionally
demanded that the charged trigger was on. 
%selection criteria were imposed on the charged trigger.
%The results of more complicated trigger systems analysis of 
%triggers do not differ from 
%$\varepsilon^{\mathrm{trig}}$ more than error.
The detection efficiency error was determined by varying selection
criteria. 
%The uncertainties of $\rho$, $\omega$, $\pi^0$ mesons parameters
%also give some contribution to the total systematic error. 
The uncertainty due to the determination
of integrated luminosity comes from the selection criteria of Bhabha
events, radiative corrections and calibration of the DC and CsI
calorimeter. The uncertainty of the model of the transition form
factor was evaluated by a difference in the detection efficiency when
simple VDM with a single $\rho$ meson as a transition particle and 
generalized VDM ($\rho$ and $\rho^{\prime}$) were used in simulation. 
The uncertainty of radiative corrections comes from the dependence on 
the emitted photon energy and the accuracy of theoretical formulae.
%The error in photon conversion probability occurs because of inexact 
%knowledge of thickness and composition of material in front of DC.
The resulting systematic uncertainty of the branching fraction quoted 
in Table~\ref{t:syserr} is 7.6\%.
\subsection{Search for $\etee$ events}
A search was performed in three main decays modes of the $\eta$ meson: 
3$\pi^0$, $\tp$ and 2$\gamma$. Selection criteria for a $\ee$-pair 
were the same as for the $\pee$ final state:
$\rho<1$~cm and $|Z_{vert}|<5$~cm; the difference between the total 
momentum of the tracks
$p=|\vec{p}_1+\vec{p}_2|$ and the momentum $p_{\gamma}$ in the
$\omega\rightarrow\eta\gamma$ process at a given energy 
$|p-p_{\gamma}|<30$~MeV/c;
the opening angle of the tracks $\Delta\psi<0.5$.
In the $\eta\to\tp$ decay mode with a four-track final state 
a $\ee$-pair was identified as a pair with opposite charges and the
smallest opening angle.
The following cuts on photons were applied:
\begin{itemize}
\item $\eta\to 3\pi^0$: the photon energy threshold was decreased 
to 30 MeV because of a higher multiplicity of soft photons in the 
final state, $N_\gamma\ge5$;
\item $\eta\to\tp$: $N_\gamma=2$,
  $M_{inv}(e^+,e^-,\gamma_1),\,M_{inv}(e^+,e^-,\gamma_2)>160$~MeV/c$^2$
  to suppress events from decays $\omtp$ followed by $\pi^0\to\ee\gamma$;
\item $\eta\to2\gamma$: $N_\gamma=2$,
the soft photon  energy $E_{\gamma,\,2}>175$~MeV to suppress QED events,
  $M_{inv}(e^+,e^-,\gamma_1),\,M_{inv}(e^+,e^-,\gamma_2)>200$~MeV/c$^2$
  to suppress events from decays $\ompg$ followed by $\pi^0\to\ee\gamma$,
  the total momentum of the tracks and photons
  $P=|\vec{p}_1+\vec{p}_2+\vec{p}_{\gamma,\,1}+\vec{p}_{\gamma,\,2}|<150$
~MeV/c.
\end{itemize}
The main background for the $\eta\to\tp$ and $\eta\to2\gamma$ decay 
modes comes from the QED processes while for the $\eta\to3\pi^0$ decay 
mode it comes from the decay $\omega\to\pi^0\pi^0\gamma$ with a 
``fake'' photon in calorimeters.
\begin{table}[tbh]
  \caption{The branching fraction, the detection efficiency at 
$s=m_\omega^2$,  the reconstruction efficiency of close tracks, 
the trigger efficiency, the number of selected events and the average 
number of background events for analyzed $\eta$ meson
    decay modes.}
  \vspace*{1mm}
  \begin{tabular}{lcccccc} \hline \hline
    Decay mode & $\mathcal{B}$, \% & $\varepsilon_{\mathrm{det}}$ &
    $\varepsilon_{\Delta\psi}$ & $\varepsilon_{\mathrm{trig}}$ & $N_{\mathrm{exp}}$ &
    $N_{\mathrm{back}}$\\ \hline
    $\eta\to$3$\pi^0$ & 32.51$\pm$0.29 & 0.070 & 0.89 & 0.98 & 0 & $<$~0.1 \\ 
    $\eta\to\pi^+\pi^-\pi^0$ & 22.6$\pm$0.4 & 0.018 & 0.89 & 0.99 & 0 & 0.2 \\ 
    $\eta\to$2$\gamma$ & 39.43$\pm$0.26 & 0.059 & 0.89 & 0.98 & 3 & 4.9 \\ \hline
  \end{tabular}
  \label{t:etamodes}
\end{table}
The result of the analysis is shown in Table~\ref{t:etamodes}.

The total number of selected events for each $\eta$ decay mode was 
expressed as:
\begin{equation}
  N = \sum_{i=1}^{n}\sigma_{BW}(s_i)\,L_i\,(1+\delta^{\eta\ee}_i)
  \,\varepsilon_{\mathrm{det}}(s_i)
  \,\varepsilon_{\Delta\psi}\,\varepsilon_{\mathrm{trig}}\,\mathcal{B}(\eta\to final).
  \label{m:ncalc}
\end{equation}
The cross section parameterization is similar to formula~(\ref{m:bw})
and includes relativistic Breit-Wigner contributions of the $\rho$ and
$\omega$ mesons, but without the $a_0$ term.
The interference model with a constant phase $\phi_{\omega-\rho}=0$ 
is used.  We used the results of our study of the process 
$\ee\to\eta\gamma$~\cite{etag} to fix
the values of the amplitude:
$\sigma^0(\rho\to\eta\ee)/\sigma^0(\omega\to\eta\ee)=0.43\pm0.09$, from which the
following relation between the branching fractions can be obtained: 
$\mathcal{B}(\rho\to\eta\ee)/\mathcal{B}(\omega\to\eta\ee)=0.65\pm0.14$.
Since the detection efficiency strongly depends on $s$, its value was 
calculated at each energy point. The
radiative corrections $\delta^{\eta\ee}$, trigger efficiency
$\varepsilon_{\mathrm{trig}}$ as well as the reconstruction efficiency
of close tracks $\varepsilon_{\Delta\psi}$ were calculated by the 
same method as for the $\pee$ events.
The average number of background events has been evaluated by MC 
simulation. The simulation of corresponding events included 
generation of ``fake'' photons~\cite{Krokthes}.
The results of all analyzed $\eta$ decay modes are taken into account 
to calculate the upper limits on the branching fractions of the $\rho$ 
and $\omega$ meson decays. The total number of
selected events has been described by summation over
expressions~(\ref{m:ncalc}) applied to each $\eta$ decay mode. To 
calculate upper limits, the Feldman-Cousins approach~\cite{errors} 
was applied with 3 events observed and 5.1 background events expected.
The final values of the upper limits were increased by 13\% in 
accordance with the systematic error.
The obtained upper limits exceed theoretical predictions by a factor of
2~\cite{Landsberg,Eidelman,Faessler}:
\begin{eqnarray}
  \mathcal{B}(\rho\to\etee) & < & 0.7\cdot10^{-5}\ (90\%\ \mathrm{C.L.}),
  \label{m:retee}\\
  \mathcal{B}(\omega\to\etee) & < & 1.1\cdot10^{-5}\ (90\%\ \mathrm{C.L.}).
  \label{m:oetee}
\end{eqnarray}
\subsection{Study of the $\omega\pi$ transition form factor}
The electromagnetic transition form factor $F(q^2)$ provides 
information on the electromagnetic structure of interacting
particles. Its dependence on squared invariant mass of virtual photon 
$q^2$ can be approximated in a small $q^2$ range as
\begin{equation}
F(q^2)=1+b\cdot q^2,
\end{equation}
where the parameter $b$ is a slope of the transition form factor.
%The $q^2$ distribution in conversion decays is analyzed in 
%work~\cite{Landsberg}. 
In experiment the information about photon 
virtuality in events of $\ompee$ decay could be taken from the 
invariant mass of the lepton pair $M_{inv}(\ee)=q^2$. Since the 
previously used selection criteria leave events with small $q^2$
only and the detection efficiency strongly depends on $M_{inv}(\ee)$, 
the cuts on total momentum, angle between the total momentum of 
tracks and photons, $M_{inv}(\ee\gamma_1)$ were omitted and
the cut on $\Delta\psi$ was increased up to 2.5 radians. To suppress
background mainly originating from $\omtp$, the kinematic fit 
requiring energy-momentum conservation and the procedure of $e/\pi$ 
separation based on the energy deposition of charged particles in 
calorimeters~\cite{phipi0ee,4pi} were additionally used.

The selected events were divided into groups of  $M_{inv}(\ee)$.
% invariant mass of $\ee$ pair.
The number of $\ompee$ events in each group was determined from the
analysis of the $\phph$ invariant mass distribution. After that
the $q^2$ dependence of the obtained number of events was fitted to
% The dependence of the number of $\ompee$ events on $q$ was
the following function from \cite{Landsberg}
\begin{eqnarray}
\frac{dN}{dq}=2q\cdot A
\cdot\frac{\alpha}{3\pi}\cdot\biggl(1-\frac{4m^2_e}{q^2}\biggr)^{1/2}
\cdot\biggl(1+\frac{2m^2_e}{q^2}\biggr)\cdot\frac{1}{q^2}\times \\ \nonumber
\times\Biggl[\biggl(1+\frac{q^2}{m^2_{\omega}-m^2_{\pi}}\biggl)^2-
\frac{4m^2_{\omega}q^2}{(m^2_{\omega}-m^2_{\pi})^2}\Biggr]^{3/2}\cdot
|F_{\omega\pi}(q^2)|^2,
\label{q2_m:5}
\end{eqnarray}
where $\alpha$ is the fine structure constant. The detector resolution 
on invariant mass $\sigma_{q}=15$~MeV/c$^2$ was taken into account. 
The floating parameters are
the normalization factor $A$ and the slope of the transition form factor
\begin{figure}[t]
\centerline{
  \psfig{file=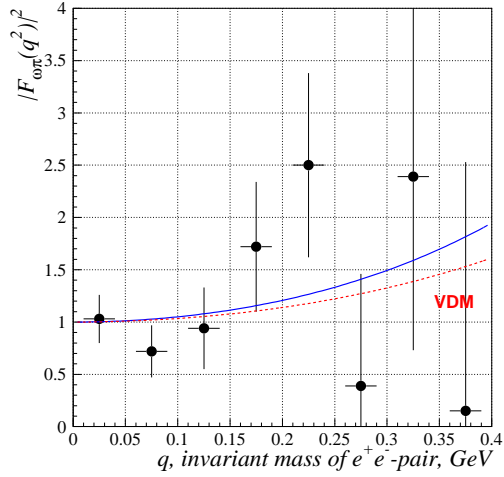,width=0.5\textwidth}
  }
%\vspace*{8pt}
\caption{\it
  Results of the fit of the data (dots with error bars) on the 
electromagnetic $\omega\pi$ transition
  form factor $|F(q^2)|^2$ (solid line). The VDM form factor is shown
  with a dotted line.\label{f:ff}}
\end{figure}
$b$. The result of the fit is shown in Fig.~\ref{f:ff}, the slope value
$b=2.5\pm3.1$~GeV$^{-2}$. The total systematic error is estimated to 
be 10\%. The main contributions to it are the $e/\pi$ separation procedure 
and selection criteria. This error is smaller than a statistical one 
for all invariant mass groups except for the group with $q<50$~MeV/c$^2$.
The obtained value agrees with the VDM prediction
$b=1/m^2_\rho=1.7$~GeV$^{-2}$ within statistical errors.

It is worth noting that modes with $\mu^+\mu^-$ in the final state
are more convenient for studies of the transition form
factor $|F_{VP}(q^2)|$~\cite{Lepton-G}.
Unfortunately, their study at CMD-2 is complicated because of the
large background coming from the dominant decay mode of the 
$\omega$ meson:
$\omega\to\tp$ and insufficiently powerful  $\mu/\pi$ separation 
at low momenta.
\section{Conclusions}
Using a data sample of 3.3~pb$^{-1}$ in the 720~--~840 MeV c.m. energy range, the
branching fraction for the $\ompee$ decay has been obtained:
\begin{displaymath}
  \mathcal{B}(\ompee)=(8.19\pm0.71\pm0.62)\cdot10^{-4}
\end{displaymath}
and the following 90\% C.L. upper limits were set:
\begin{eqnarray*}  
  \mathcal{B}(\rhpee) & < & 1.6\cdot10^{-5}, \\
  \mathcal{B}(\rho\to\etee) & < & 0.7\cdot10^{-5}, \\
  \mathcal{B}(\omega\to\etee) & < & 1.1\cdot10^{-5}.
\end{eqnarray*}

{\bf Acknowledgment}

The authors are grateful to the staff of VEPP-2M for excellent 
performance of the collider, and to all engineers and technicians 
who participated in the design, commissioning and operation of CMD-2.
This work is supported in part by grants DOE DEFG0291ER40646, NSF PHY-0100468,
PST.CLG.980342, RFBR-03-02-16280-a, RFBR-03-02-16477,
RFBR-03-02-16843, RFBR-04-02-16217, RFBR-04-02-16233-a, and RFBR-04-02-16434.

\end{document}